\documentclass[amsmath,notitlepage,twocolumn,prb,floatfix,aps,10pt]{revtex4-1}
\RequirePackage{varioref}
\RequirePackage{amssymb}
\RequirePackage{bm}
\RequirePackage{graphicx}
\RequirePackage{color}

\newcommand{\mathcommand}[3][0]{\newcommand{#2}[#1]{\ensuremath{#3}}}
\renewcommand{\vec}[1]{\ensuremath{\text{\textbf{#1}}}} 
\newcommand{\be}{\begin{equation}}
\newcommand{\ee}{\end{equation}}

\newcommand{\ts}[2]{{#1}_{\textnormal{#2}}} 

\newcommand{\Tr}[2][]{\text{Tr}_\text{#1}\left\{#2\right\}}

\mathcommand[1]{\smallket}{|#1\rangle}
\mathcommand[1]{\smallbra}{\langle #1|}
\mathcommand[1]{\bigket}{\bigl|#1\bigr\rangle}
\mathcommand[1]{\bigbra}{\bigl\langle #1\bigr|}
\mathcommand[1]{\biggket}{\biggl|#1\biggr\rangle}
\mathcommand[1]{\biggbra}{\biggl\langle #1\biggr|}
\mathcommand[1]{\ket}{\left| #1 \right\rangle}  
\mathcommand[1]{\lket}{\bigl| #1 \bigr)}  
\mathcommand[1]{\bra}{\left\langle #1 \right|}  
\mathcommand[1]{\lbra}{\bigl( #1 \bigr|}  
\newcommand{\ketbra}[2]{\left| #1 \right\rangle\!\!\left\langle #2
\right|} 

\mathcommand[1]{\bbra}{\biggl\langle #1 \biggr|}
\mathcommand[1]{\bket}{\biggl| #1 \biggr\rangle}

\newcommand{\refeq}[1]{Eq.~\eqref{#1}}
\newcommand{\refsec}[1]{Sec.~\ref{#1}}
\newcommand{\reffig}[1]{Fig.~\ref{#1}}

\newcommand{\redmat}[3]{\bra{#1}\!|#2|\!\ket{#3}}
\newcommand{\bigredmat}[3]{\bigbra{#1}\!\bigl|#2\bigr|\!\bigket{#3}}
\newcommand{\vecirrep}[1]{\hat{\vec{#1}}}
\newcommand{\sixj}[6]{
\left\{\!\begin{array}{lcr}#1&#2&#3\\
                           #4&#5&#6\end{array}\!\right\}}

\begin{document}
\title{Gauge-invariant Adiabatic Two-Qubit Gates for Exchange-Only Qubits}
\author{Thaddeus D. Ladd and Cody Jones}
\affiliation{HRL Laboratories, LLC,
3011 Malibu Canyon Rd., Malibu, CA 90265}
\date{\today}
\begin{abstract}
We extend recent work on a leakage-protected, adiabatic entangling gate for exchange-only spin qubits [Doherty and Wardrop, PRL \textbf{111}, 050503 (2013)] by adapting to a setting where single spins are not assumed to be polarized on preparation.  Previous gate constructions do not function correctly when ``gauge spins'' are uninitialized, because the entangling gate has different, non-trivial action in different gauge subspaces.  Our construction inherits many of the desirable features of the previous work while addressing the gauge-dependent behavior.  Using numerical simulation, we show that the resulting gate implements the same logical operation in both gauge subspaces to first order in perturbation theory, and second-order terms introduce an error that decreases quadratically in the duration of the gate.  We add $1/f$ charge noise to voltages modulating exchange in this model, which introduces errors that increase with gate time, to show that there is an optimal gate duration for a given set of device parameters.
\end{abstract}
\maketitle

\section{Introduction}
Of the many types of qubits under development for quantum computation\cite{Ladd2010}, semiconductor exchange-only spin qubits benefit from being defined and controlled only by voltages on gates in a platform resembling classical microelectronics\cite{DiVincenzo2000}.  The ability to do exchange-only control, however, comes at the cost of complexity of encoding and the possibility of qubits leaking into unencoded spaces.

Two-qubit gates are especially challenging with encoded universality.  Recently, an adiabatic two-qubit gate was proposed\cite{Doherty2013} in the context of Resonant eXchange (RX) qubits\cite{Medford2013}, and proposed for use with Always-on Exchange-ONly (AEON) qubits\cite{Shim2016}. This form of two-qubit gate has both advantages and disadvantages relative to pairwise-exchange pulse sequences\cite{DiVincenzo2000,Fong2011}; one disadvantage relative to recent pulsed gates\cite{Fong2011} is that, as published, the gate does not preserve the gauge freedom of the 3-spin encoding,  requiring instead some means for spin polarization.  In the present article, we review this gate and introduce a mathematical formalism employing the Wigner-Eckart theorem to elucidate its structure.  This formalism is beneficial for not only understanding the algebra behind the operation of this gate, but also reveals a small modification that allows this two-qubit gate to preserve gauge-freedom and therefore remove requirements for spin polarization.

In what follows, we first define, review, and discuss encoded exchange-only qubits in general in \refsec{sec:DFSdef}, introducing the Wigner-Eckart theorem to calculate relevant matrix elements.  In \refsec{sec:general_adiabatic} we then give a general introduction to the adiabatic two-qubit gate earlier introduced in Ref. \onlinecite{Doherty2013}, and show in \refsec{sec:invariant_gate} how to modify this gate to preserve the gauge freedom of triple-dot exchange-only qubits.  The general presentation is made more specific with an example qubit choice, geometrical layout, and numeric matrix evaluations in \refsec{sec:examples}, and we then briefly examine the effect of noise and higher-order corrections to the gate in \refsec{sec:errors}, especially considering new noise terms relative to those handled in Ref. \onlinecite{Doherty2013} when preserving gauge freedom.

\section{DFS Qubits and Two-qubit Gates}
\label{sec:DFSdef}
\begin{figure}[b]
\includegraphics[width=0.7\columnwidth]{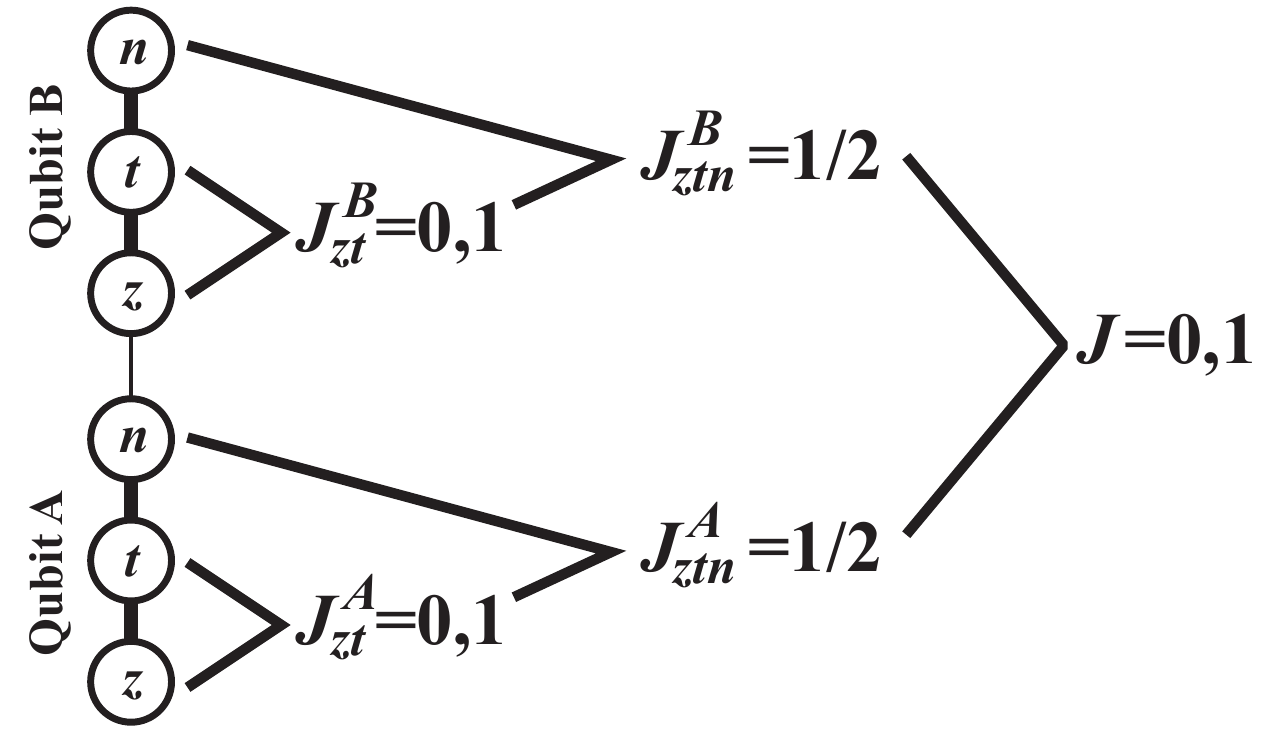}
\caption{Addition of angular momentum for a pair of DFS qubits.  Six spins, each with $S=1/2$, combine first via their $zt$ pairs into the qubit quantum numbers $J^A_{zt}$ and $J^B_{zt}$.  These each combine with $n$ spins to form the $J^{A,B}_{ztn}$ quantum numbers, maintained as $1/2$ for the qubit encoding.  The two qubits then combine to form a total angular momentum $J$, which may be 0 or 1.}
\label{Jfig}
\end{figure}

The ability to operate spin qubits using only Heisenberg exchange depends critically on the encoding into Decoherence Free Subsystems (DFS).  A decoherence-free subsystem is an encoding in which one or more collective quantum numbers of the composite system may be ignored, such that noise that couples to those quantum numbers does not cause decoherence.  If there is multiplicity of equivalent subspaces, we refer to the system as a subsystem and the untracked degrees of freedom as ``gauge" freedom.  If there is only a single encoded subspace, the term ``decoherence free subspace" is used.  In the context of spins, the gauge degree of freedom is the projection of total angular momentum, $m$, which couples to any common magnetic field across those spins.  Common magnetic field noise, therefore, is rejected.  While the common-mode rejection feature of the DFS lends it its name, the importance for its use in the present context is to provide an encoding in which Heisenberg-exchange provides encoded universality.

The DFS for three spins, arbitrarily labeled $z,t$ and $n$, has quantum numbers $J_{zt}$, $J$, and $m$, where $J_{zt}$ is the total angular momentum of spins $z$ and $t$ while $J$ is the total angular momentum of all spins.  The DFS is the subsystem manifold $J=1/2$.  The $J_{zt}$ number, however, takes on values $0$ and $1$ and may be altered by exchange, providing the qubit.  Note that not all exchange-only qubit encodings directly employ the $J_{zt}$ quantum number; in general our qubit states may be notated in terms of the $\ket{J_{zt},J,m}$ angular momentum states as
\begin{align}
\ket{0;m} &= \sum_{J_{zt}=0,1} W_{0,J_{zt}}\ket{J_{zt},1/2,m},\\
\ket{1;m} &= \sum_{J_{zt}=0,1} W_{1,J_{zt}}\ket{J_{zt},1/2,m}.
\end{align}
For example, the RX\cite{Medford2013} and AEON\cite{Shim2016} qubits convert via the rotation $W_{00}=-1/2, W_{11}=1/2, W_{01}=W_{10}=-\sqrt{3}/2$, which results from a SWAP of spins $t$ and $n$.  We refer to a DFS qubit as having $W$ given by the identity matrix.  Regardless of the chosen logical basis, Heisenberg exchange between these three spins conserves $J$, leaving the DFS intact; it also conserves the quantum number $m$ and is therefore naturally gauge-invariant.
We also have a single leaked state for each gauge value, notated as
\be
\ket{Q;m} = \ket{1,3/2,m}.
\ee
The decoherence free subspace for four spins is similar; in this case we have quantum numbers $J_{zt},J_{ztn},J$ and $m$, and the encoding is for $J_{ztn}=1/2,J=0$.  This is effectively the same as the three-spin DFS where the gauge spin $m$ forms a singlet with the fourth spin; for this system, gauge-invariant sequences for the three-spin DFS may be applied to the four-spin DFS as well.  In this case there are nine leaked states with $J=1$ and 4 with $J=2$.

The action of exchange between any two spins may be understood through the Wigner-Eckart theorem, noting that each spin vector operator $\vecirrep{S}_k$ is an irreducible tensor operator of order 1.  As this theorem provides important insight for what follows, it is important to introduce it now for its notation and computational machinery. Following the derivation and notation of Messiah\cite{Messiah}, this theorem says
\begin{multline}
\bra{\tau_1\tau_2 J_1 J_2 J m}\vec{S}_j\cdot \vec{S}_k\ket{\tau_1'\tau_2'J_1'J_2'J'm'}
=\\
\delta_{JJ'}\delta_{mm'}(-1)^{J+J_2+J_1'}
\sixj{J_1}{1}{J_1'}{J_2'}{J}{J_2}
\times\\
                \redmat{\tau_1J_1}{\vecirrep{S}_j}{\tau_1'J_1'}
                \redmat{\tau_2J_2}{\vecirrep{S}_k}{\tau_2'J_2'}.
\label{exchangeformula}
\end{multline}
which introduces the Wigner 6$j$ symbols in curly braces and the reduced matrix elements $\redmat{\cdot}{\vecirrep{S}}{\cdot}$.  Here the variables $\tau_j$ will capture internal quantum numbers, which in our context are more $J$ quantum numbers for subsets of spins. For the 3-spin DFS qubit, we may notate the encoded states as $\ket{J_{zt},J=1/2,m}$. First, the fact that exchange conserves $J$ and $m$ is a trivial consequence of this theorem.  Second, if two spins are grouped together with a quantum number, such as the $z$ and $t$ spins in our DFS qubit, then the action of exchange is simply a phase-shift; i.e.
\begin{multline}
\label{Jzt}
\bra{J_{zt},J,m}\vec{S}_{z}\cdot\vec{S}_t\ket{J_{zt},J,m} = (-1)^{J_{zt}+1}
\times\\
\sixj{1/2}{1}{1/2}{1/2}{J_{zt}}{1/2}
\redmat{1/2}{\vecirrep{S}}{1/2}^2 = \frac{1}{4}-\delta_{J_{zt}0}.
\end{multline}
Since this operation is diagonal, it is likened to a qubit $z$ rotation on the Bloch sphere defined by qubit states $\ket{J_{zt}=0,J=1/2,m}$ and $\ket{J_{zt}=1,J=1/2,m}$ (motivating the choice of name $z$ for the first spin).  More explicitly, introducing Pauli operators in terms of our encoded gauge-qubit states as $\ket{0;m}$ and $\ket{1;m}$ as $X=\sum_m \ketbra{0;m}{1;m}+\ketbra{1;m}{0;m}$ and likewise for $Y,Z$ and $I$, we find that for the DFS qubit $\vec{S}_{z}\cdot\vec{S}_t \rightarrow 1/4-(I+Z)/2$.
A less trivial use of the Wigner-Eckart theorem is the Heisenberg exchange between spins $t$ and $n$, which reads
\begin{multline}
\bra{J_{zt},J,m}\vec{S}_{t}\cdot\vec{S}_n\ket{J_{zt}',J,m}
=(-1)^{J+1/2+J_{zt}'}\times\\
\sixj{J_{zt}}{1}{J_{zt}'}{1/2}{J}{1/2}
                \redmat{J_{zt}}{\vecirrep{S}_t}{J_{zt}'}
                \redmat{1/2}{\vecirrep{S}_n}{1/2}
\\
\rightarrow\frac{1}{4}-\frac{I}{2}-\cos\left(\frac{2\pi}{3}\right)\frac{Z}{2}
             +\sin\left(\frac{2\pi}{3}\right)\frac{X}{2}.
\end{multline}
Hence for the DFS qubit this matrix describes a rotation about an axis tipped $2\pi/3$ away from $z$ on the $xz$ plane for the encoded Bloch sphere.  (This is sometimes referred to as the $n$-axis motivating the naming convention for the third spin).

We note briefly that if collinear inhomogeneous magnetic fields are introduced into the 3-spin DFS (that is, varying magnetic field at dots $z$,$t$, and $n$ but whose vector direction is the same for each), the $J$-quantum number is no longer conserved.  Collinear magnetic gradients cause leakage from the encoded $J=1/2$ states to the leaked $J=3/2$ states.  The $m$-quantum number remains conserved, but the rotations in SU(3) caused by inhomogeneous magnetic fields are in opposite directions for different $m$-states, breaking gauge-invariance.

Neglecting the possibility of inhomogeneous magnetic fields for now, we now examine two-qubit gates and note that there remains a gauge degree of freedom, $m$, now in place for all spins, and exchange-only action will retain gauge-invariance relative to this quantum number.  However, if we examine the quantum numbers of this system, we must now consider another type of invariance, namely $J$-invariance.  Consider two 3-spin DFS qubits, as depicted in \reffig{Jfig}, where we still label spins as $z$, $t$, $n$ but now use a superscript to indicate whether each spin is in qubit $A$ or $B$, so qubit $A$ has $J$ quantum numbers $J_{zt}^A, J_{ztn}^A$ and qubit $B$ has $J$ quantum numbers $J_{zt}^B,J_{ztn}^B$.  The combined system, however, has total angular momentum $\vec{J}=\vec{J}^A+\vec{J}^B$.  For the two qubits in DFS with $J^{A,B}_{ztn}=1/2,$ the total angular momentum of the pair of spins may be $J=0$ or $J=1$.  A $J$-invariant two-qubit quantum gate is one which has the same action on the encoded qubit quantum numbers $J_{zt}^{A,B}$ for both the $J=0$ and $J=1$ subspaces.  Note that the $J=0$ subspace is 5-dimensional, while the $J=1$ subspace is 9-dimensional.  Since exchange conserves $J$, these subspaces cannot be mixed by exchange, and neglecting overall phase our control space is effectively SU(5)$\times$SU(9).

We use the phrase $J$-invariance because it is not precisely equivalent to gauge-invariance; within $J=1$, the $m=0,\pm 1$ quantum number remains a good quantum number for exchange-only operation and, like the $m=\pm 1/2$ number for a single-qubit, may be ignored in the absence of inhomogeneous magnetic fields.  However, the requirement of $J$-invariance for two-qubits is a consequence of \emph{gauge freedom} of single qubits: the $J=0$ subspace of two qubits results from the single-qubit gauges forming a singlet, while the $J=1$ subspace of two qubits results from the single-qubit gauges forming a triplet.  In this sense $J$-invariance is an extension of gauge-invariant operation for multi-qubit DFS systems.

One effectively escapes the complication of $J$-invariance using the 4-spin decoherence-free subsystem, which has no single-qubit gauge freedom.  In this case for spins $z,t,n$ and $c$ in each qubit we have the quantum numbers $J_{zt}^A,J_{ztn}^A,J_{zt}^B,J_{ztn}^B,J_{ztnc}^A,J_{ztnc}^B,J,m$.  Our encoded subspace enforces most of these quantum numbers to be fixed values, namely $J_{ztn}^A=J_{ztn}^B=1/2$ and $J_{ztnc}^A=J_{ztnc}^B=J=m=0$, and a two-qubit gate need only enforce appropriate action on the encoded qubit quantum numbers $J_{zt}^A$ and $J_{zt}^B$ within this subspace.  However, to accomplish a quantum gate, we will use exchange operations between $A$-spins and $B$-spins, which fail to conserve the encoding and in general introduce a 5+9=14 dimensional subspace.  As the SU(14) control subspace here is larger than the SU(5)$\times$SU(9) subspace of the six-spin case, pulse-sequence design is generally harder in this system, and it may be that the most efficient two-qubit gates of the 4-spin-DFS result from $J$-invariant 3-spin DFS gates.

For two-qubit gates composed of sequences of sequential, pairwise exchange between spins, there are several known fully entangling two-qubit gates.  The CNOT-like sequence first published in 2000 by DiVincenzo et al.\cite{DiVincenzo2000} was not $J$-invariant; this pulse sequence only has the correct action in the $J=1$ subspace.  Note that DFS qubits may be physically restricted to this subspace by eliminating gauge freedom and assuring that the $m$ quantum number of every qubit is identical, which requires some mechanism for spin polarization.  Spin polarization is possible with high magnetic fields and low temperature (i.e. when $g \mu B \gg k_B T$).  However, high magnetic fields can interfere with singlet preparation, which is also needed for the DFS qubit, if the energy of the triplet $T_{--} = \ket{\downarrow \downarrow}$ dips below that of singlet; this happens when the Zeeman splitting is larger than the energy gap between ground and excited spatial wavefunctions for a doubly charged quantum dot.  For example, in silicon it is preferable to keep Zeeman splittings smaller than valley splittings\cite{Eng2015}.

Fortunately, a $J$-invariant two-qubit gate was discovered by a genetic algorithm in 2013 by Fong and Wandzura\cite{Fong2011}.  This pulse sequence uses an inner core of 14 square-roots-of-SWAP between spins, plus SWAPs to adjust to any particular layout of spins, with 20-pulses for a controlled-Z (CZ) gate between qubits whose $z$-type spins are coupled.  Searches for shorter sequences in arbitrary layouts have yielded no sequence shorter than this 14-pulse core sequence, although different spin-coupling layouts may use SWAPs with varying degrees of efficiency\cite{Setiawan2014}.  For the 4-spin DFS, sequentially pulsed two-qubit gates are known: a multi-exchange construction was presented by Bacon {\it et al.} in 2000\cite{Bacon2000} and, being a $J$-invariant gate, the single-exchange Fong-Wandzura sequence may also be used for this encoding\cite{Fong2011}.  For an analytic derivation of the Fong-Wandzura sequence, see Ref.~\onlinecite{Zeuch2016}.

\section{Adiabatic Two-Qubit Gates}
\label{sec:general_adiabatic}
\begin{figure}[th]
\includegraphics[width=\columnwidth]{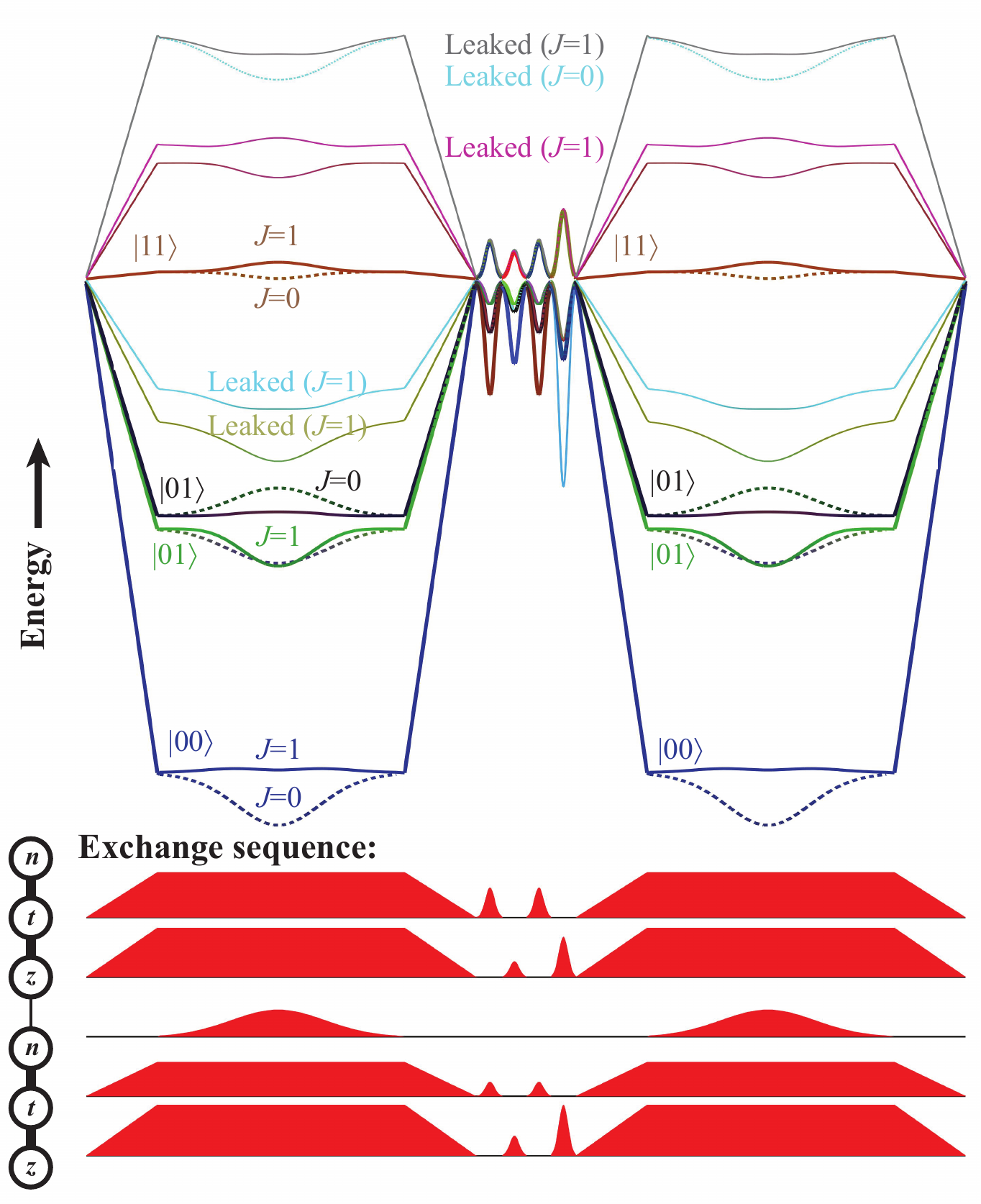}
\caption{Schematic of the $J$-invariant adiabatic two-qubit gate.  (a) Example energy levels, which are first ramped to single-qubit bias levels and then modulated via a Gaussian pulse.  The corresponding exchange sequence is shown in (b).  Each row in (b) corresponds to the value of pairwise exchange between spin pairs as indicated by the linear array of spins shown on the left, following \reffig{Jfig}.  In the energy diagram, the 9 energy levels of $J=1$ are shown as solid lines; four of these are the encoded states, here labeled $\ket{00},\ket{01},\ket{10}$ and $\ket{11}$, and the rest are leaked states.  The 5 energy levels of $J=0$ are shown as dashed lines, which track the solid lines for the single-qubit biases but not for the two-qubit pulse.  The case $m=0$ is shown, for which the single-qubit $J=0$ and $J=1$ energies are identical; not shown are the $m=\pm 1$ cases, which in uniform finite magnetic field would be exact copies of the $m=0$ spectrum split off by the electron Zeeman energy.  Two adiabatic evolutions are shown, each intended to perform a $\sqrt{CZ}$ pulse in addition to single-qubit $Z$-rotations,  interspersed with a refocusing single-qubit composite-pulse sequence.   The 4-pulse sequence shown corresponds to the $\pi$-pulse about the encoded $y$-axis of the DFS qubit, referred to as the $\ts{Y}{echo}$ pulse~\cite{Eng2015}.}
\label{pulse_schematic}
\end{figure}
In contrast to entangling gates using a sequence of pulses, the adiabatic DFS-entangling gates introduced by Doherty and Wardrop in 2013\cite{Doherty2013} are appealing because they can be implemented as a ``single-shot'' operation on all spins in two DFS blocks.  Focusing on three-spin DFS qubits, these gates behave as follows.  First, at some time prior to the gate (and possibly substantially before), ``large" exchange values are introduced between the three spins within each DFS qubit.  The purpose of these large exchange values for our consideration here is to lift the degeneracy between the $J=1/2$ single-qubit encoded states and the $J=3/2$ leaked states.  Note that this cannot be done with a single pairwise exchange; for example \refeq{Jzt} leaves the $\ket{J_{zt}=1,J=1/2,m}$ encoded state degenerate with the $\ket{J_{zt}=1,J=3/2,m}$ leakage state.  At least two exchange terms must be nonzero.  The Hamiltonian to be activated for either qubit may in general be written
\begin{multline}
H_1^A = \Omega^A_{zt}\vec{S}_z^A\cdot\vec{S}_t^A +
                 \Omega^A_{tn}\vec{S}_t^A\cdot\vec{S}_n^A +
                 \Omega^A_{zn}\vec{S}_z^A\cdot\vec{S}_n^A,
\end{multline}
for positive exchange rates $\Omega^A_k$, $k=zt,tn,zn$.  Omitting the superscript for simplicity, the eigenenergies of the $J=1/2$ states relative to the leaked $J=3/2$ states are
\be
\xi_\pm = \frac{1}{2}\biggl[-\sum_{k} \Omega_{k}\pm \sqrt{\sum_{k} \Omega_k^2-\sum_{k < \ell} \Omega_k\Omega_\ell}\biggr],
\ee
from which it may readily be seen that the $+$ eigenvalue vanishes if just one of the exchange amplitudes $\Omega_k$ is nonzero.  These eigenstates are of course superpositions of the encoded angular momentum states or the encoded qubit states
\begin{multline}
\ket{\xi_\pm,m}=\sum_{J_{zt}=0,1} V_{\pm,J_{zt}}\ket{J_{zt},J=1/2,m} \\
 = \sum_{J_{zt}=0,1}
V_{\pm,J_{zt}} [W^*_{0,J_{zt}} \ket{0;m}+W^*_{0,J_{zt}} \ket{1;m}].
\label{xi_def}
\end{multline}
We will define a unitary $R_A$ as the rotation between the diagonal basis of $H_1^A$ and the desired qubit computational basis from \refsec{sec:DFSdef}, which is $W V^\dag$ in the encoded logical subspaces and unity for the $J_{ztn}^A=3/2$ leakage states.

The particular choice of exchange values $\Omega^A_k$ may be chosen according to a variety of design choices.  The RX qubit\cite{Medford2013} and the AEON qubit\cite{Shim2016} choose these values to target a particular sweet spot in charge noise according to Fermi-Hubbard-based models for exchange; for a discussion of these and other triple-dot qubit models, see Ref.~\onlinecite{Russ2017}.  In practice with quantum dots, these $\Omega_k^A$ and $\Omega_k^B$ values may require tuning relative to constraints imposed by dot disorder and capacitance properties.  For the present derivation, we leave the exact values or ratios of these parameters as unconstrained, except insofar as at least two of them are made sufficiently positive so as to sufficiently split the energies of encoded subspaces relative to leaked subspaces.

After applying $H_1^A$ and $H_1^B$ to the individual qubits, adiabatic two-qubit exchange-only gates then introduce one of the interqubit exchange parameters.  If we couple spin $a$ from qubit $A$ and spin $b$ from qubit $B$, our two-qubit hamiltonian term may be written
\be
H_2^{AB}(t) = \Omega^{AB}_{ab}(t)\vec{S}_a^A\cdot\vec{S}_b^B.
\ee
Here $\Omega^{AB}_{ab}(t)$ is slowly ramped on and then off again, and $\Omega^{AB}_{zz}(t)$ is kept sufficiently small to assure $\Omega^{AB}_{zz} \ll \xi_\pm^A,\xi_\pm^B$; i.e. its total amplitude is kept smaller than any of the encoded-to-leaked splittings of either qubit.  Under these assumptions we may assume adiabatic phase accrual proportional to the pulse area
\be
\Phi^{AB} = \int_0^T \Omega^{AB}_{ab}(t) dt.
\ee
To see that a two-qubit gate results, we suppose that each encoded, biased eigenstate $\xi_s^{A,B}$ for qubit $A$ or $B$ and eigenvalue $s=\pm$ is altered via first-order perturbation theory by the first-order difference energy $\bra{\xi_s^A,\xi_r^B}H_2^{AB}(t)\ket{\xi_s^A,\xi_r^B}$, and for the moment we assume no degeneracy between qubit states.  Each eigenstate develops a phase during the pulse of duration $T$ of
\be
\phi_{sr} = \xi_{s}^AT+\xi_{r}^BT+\bra{\xi_s^A,\xi_r^B}\vec{S}_a^A\cdot\vec{S}_b^B\ket{\xi_s^A,\xi_r^B}\Phi^{AB}.
\ee
This phase rotation produces evolution (within the encoded states only) of the form
\begin{multline}
\label{Udef}
U =
[R^A\otimes R^B]\cdot\\
    \exp[-i(c_{00} + c_{10} Z^A + c_{01}Z^B + c_{11}Z^AZ^B)]\cdot\\ [R^A\otimes R^B]^\dag,
\end{multline}
where we recall that $R^A$ was defined above as single-qubit rotations which map the computational basis into the diagonal basis of the entangling gate; an example will be given in \refsec{sec:examples}.  The $c_{ij}$ coefficients are linear combinations of the phases accumulated by the eigenstates in \refeq{xi_def}:
\begin{align*}
c_{00} &= (\phi_{++}+\phi_{--}+\phi_{+-}+\phi_{-+})/4,\\
c_{01} &= (\phi_{++}-\phi_{--}+\phi_{+-}-\phi_{-+})/4,\\
c_{10} &= (\phi_{++}-\phi_{--}-\phi_{+-}+\phi_{-+})/4,\\
c_{11} &= (\phi_{++}+\phi_{--}-\phi_{+-}-\phi_{-+})/4.
\end{align*}
and if the timing is arranged so that $c_{11}=(2m+1)\pi/4$ for any integer $m$ then the resulting gate will be equivalent to a CZ, up to single-qubit phase corrections.

If there is degeneracy between the two qubits, then adabiatic evolution includes coherent population transfer between the degenerate states.  (Phases accrue on those states diagonal in $H_2^{AB}(t)$).  This would of course be especially problematic if a leakage state in one qubit is degenerate with an encoded logical state of another, which is possible for certain pathological choices of biases $\Omega^X_k$.  For more reasonable bias choices, however, it is the encoded states which may be degenerate, possibly by design.  In this case, the resulting gate will involve some component of qubit swapping in addition to other phases, as has been discussed in Refs. \onlinecite{Doherty2013} and \onlinecite{Wardrop2016}.  Depending on which spins are coupled, any combination of CZ and SWAP may occur.  This still allows entangling gates, but for the present discussion we will focus on the scenario where the qubit biases are non-degenerate, and the qubit separation energies all exceed the peak value of $\Omega^{AB}_{ab}(t)$ by an amount to be quantified in \refsec{sec:errors}.

Besides the possibility of degeneracy causing unintended evolution, there are of course errors due to higher-order terms, non-adiabaticity, charge noise, and magnetic noise.  We shall address these more as well in \refsec{sec:errors}, although most of these errors, as well as several examples of qubit connectivities, were discussed in Doherty and Wardrop\cite{Doherty2013,Wardrop2016}.  What those articles did not address is the issue of gauge freedom and $J$-invariance, instead deriving the action of the gate only in the $J=1$ (spin-polarized subspace).  We now address the fact that this form of gate may naturally be made $J$-invariant with minimal extra overhead.

\section{$J$-Invariant Adiabatic Two-Qubit Gates}
\label{sec:invariant_gate}

To see how $J$-invariance can be enforced, let us look at our perturbative phases in general, supposing that our two-qubit exchange of choice couples spin $a$ of qubit $A$ to spin $b$ of qubit $B$.  These are given by
\begin{widetext}
\begin{multline}
\phi_{sr} = \xi_{s}^AT+\xi_{r}^BT+\Phi^{AB}
    \sum_{J_{zt}^A,{J_{zt}^A}'}\sum_{J_{zt}^B,{J_{zt}^B}'}
        V_{s,J_{zt}^A}^{A*}V_{s,{J_{zt}^A}'}^A
        V_{r,J_{zt}^B}^{B*}V_{s,{J_{zt}^B}'}^B
    \\
    \bigbra{J_{zt}^A,J_{ztn}^A,J_{zt}^B,J_{ztn}^B,J,m}
        \vec{S}_a\cdot\vec{S}_b
    \bigket{{J_{zt}^A}',J_{ztn}^A,{J_{zt}^B}',J_{ztn}^B,J,m},
\end{multline}
in which we maintain the encoding condition $J_{ztn}^A=J_{ztn}^B=1/2$. We now exploit the Wigner-Eckart theorem, \refeq{exchangeformula}, to note the $J$-dependent structure of this matrix element, i.e.
\begin{multline}
    \bigbra{J_{zt}^A,J_{ztn}^A,J_{zt}^B,J_{ztn}^B,J,m}
        \vec{S}_a\cdot\vec{S}_b
    \bigket{{J_{zt}^A}',J_{ztn}^A,{J_{zt}^B}',J_{ztn}^B,J,m} =
\\
    (-1)^{J+1}
\sixj{J_{ztn}^A}{1}{J_{ztn}^A}{J_{ztn}^B}{J}{J_{ztn}^B}
                \bigredmat{J_{zt}^A,J_{ztn}^A}{\vecirrep{S}_a}{{J_{zt}^A}',J_{ztn}^A}
                \bigredmat{J_{zt}^B,J_{ztn}^B}{\vecirrep{S}_b}{{J_{zt}^B}',J_{ztn}^B}.
\end{multline}
The reduced matrix elements in this expression will depend on the specific choices of spins $a$ and $b$, and we will consider some examples below.  The critical observation, however, is that these matrix elements do not depend on $J$.  In fact, we may evaluate the Wigner 6$j$ symbol in the $J_{ztn}^A=J_{ztn}^B=1/2$ subspace (using, for example, Eq.~9.5.2.7 of Ref.~\onlinecite{VMK}) to
\be
    \bigbra{J_{zt}^A,1/2,J_{zt}^B,1/2,J,m}
        \vec{S}_a\cdot\vec{S}_b
    \bigket{{J_{zt}^A}',1/2,{J_{zt}^B}',1/2,J,m} =
    \frac{(-1)^{J+1}}{(2+J)!}
                \bigredmat{J_{zt}^A,1/2}{\vecirrep{S}_a}{{J_{zt}^A}',1/2}
                \bigredmat{J_{zt}^B,1/2}{\vecirrep{S}_b}{{J_{zt}^B}',1/2},
\ee
for $J=0,1$.
\end{widetext}

Therefore, we can simplify our phase expression to
\be
\phi_{sr}=\xi_{s}^AT+\xi_r^BT+\frac{(-1)^{J+1}}{(2+J)!}\Phi^{AB} \Lambda_s^A \Lambda_r^B,
\ee
where detailed structure factors are (omitting single-qubit superscripts)
\be
\label{structurecalc}
\Lambda_{s}^A =
    \sum_{J_{zt}^A,{J_{zt}^A}'}
        V_{s,J_{zt}^A}^{A*}V_{s,{J_{zt}^A}'}^A
            \bigredmat{J_{zt}^A,1/2}{\vecirrep{S}_a}{{J_{zt}^A}',1/2},
\ee
and likewise for $\Lambda_{r}^B$.  The reduced matrix elements (dropping superscripts for simplicity) in the $J_{zt}=0,1$ basis are\cite{Messiah}
\be
\label{matrixcalc}
\bigredmat{J_{zt},1/2}{\vecirrep{S}_{z,t}}{J_{zt}',1/2} =
\begin{pmatrix}
0 & \mp 1/\sqrt{2} \\
\mp 1/\sqrt{2} & \sqrt{2/3}
\end{pmatrix}
\ee
where the off-diagonal elements are negative for the $z$ spin and positive for the $t$ spin, and
\be
\bigredmat{J_{zt},1/2}{\vecirrep{S}_n}{J_{zt}',1/2} = \delta_{J_{zt},J_{zt}'}
\frac{(-1)^{J_{zt}}\sqrt{6}}{(2+J_{zt})!}.
\ee

To make a maximally entangling gate, then, we could simply require
\begin{multline}
c_{11} = \frac{(-1)^{J+1}}{4(2+J)!}\Phi^{AB}[
 \Lambda_{+}^A\Lambda_{+}^B
+\Lambda_{-}^A\Lambda_{-}^B
-\Lambda_{+}^A\Lambda_{-}^B
-\Lambda_{-}^A\Lambda_{+}^B]
\\
=(2m+1)\pi/4,
\end{multline}
for some integer $m$.  This condition is met if we assure
\be
\Phi^{AB}[
 \Lambda_{+}^A\Lambda_{+}^B
+\Lambda_{-}^A\Lambda_{-}^B
-\Lambda_{+}^A\Lambda_{-}^B
-\Lambda_{-}^A\Lambda_{+}^B]=6\pi.
\ee
The complication we now encounter, however, involves the single-qubit corrections, which cannot necessarily be made using exchange-only single-qubit operations if they vary with $J$ or with single-qubit gauge.  The $c_{01}$ and $c_{10}$ terms will, in general, include terms varying with $J$, and in general the two $J$ subspaces will receive different single-qubit $Z$-rotations in the $\xi_{\pm}^{A,B}$ eigenstate basis.  To remedy this problem, we use the commutative nature of our assumed adiabatic evolution and require the construction of the single-qubit operations
\begin{multline}
\label{Pidef}
\Pi^A = \sum_{m} -i e^{i(\eta-\eta')}\ketbra{\xi_+^A,m}{\xi_-^A,m}
\\
       -ie^{i(\eta+\eta')}\ketbra{\xi_-^A,m}{\xi_+^A,m}
       +e^{i\xi}\ketbra{Q;m}{Q;m}
\end{multline}
for some arbitrary phases $\eta$ and $\eta'$, and likewise for qubit $B$.  The specific implementation of this gate depends on the choice of bias fields $\Omega_k^A$; however since these are single-qubit operations, we may easily in all cases calculate how to achieve arbitrary single-qubit evolution, and since we maintain full gauge-freedom for single-qubit operations, this operation is guaranteed to be $J$-invariant.  Note that Ref.~\onlinecite{Wardrop2016} proposed an echo-pulse to suppress noise on exchange control, but did not consider its utility for a $J$-invariant gate.

One bias-independent way to construct a gate of the form \refeq{Pidef} is to establish a $\pi$-rotation about the $Y$-axis in the Bloch sphere of the DFS qubit defined by the quantum numbers $J_{zt}^A$ and $J_{zt}^B$.  Since all single-qubit DFS couplings are in the $xz$-plane in this encoding (a consequence of the Condon-Shortley convention for Clebsch-Gordan coefficients\cite{VMK}), a $Y$-pulse will refocus any bias.  Such a $Y$-pulse is depicted in the pulse schematic of \reffig{pulse_schematic}.  A $Y$-echo pulse to refocus low-frequency noise on exchange evolution using four pulses of pairwise exchange has been demonstrated in an isotopically enhanced triple-dot in Si/SiGe using pairwise exchange in Ref.~\onlinecite{Eng2015}.  Of course other pulsing strategies such as for the RX-qubit\cite{Medford2013} or AEON-qubit pulsing\cite{Shim2016} may be alternatively be employed for the same task.

Therefore our full $J$-invariant two-qubit gate, as depicted in \reffig{pulse_schematic}, may be constructed via the composite construction (referring back to \refeq{Udef}, and only considering the encoded subspace up to overall phase) as
\begin{multline}
\ts{U}{composite} = U\cdot \Pi^A\Pi^B\cdot U
\\
=
e^{-2ic_{00}} [R^A\otimes R^B]
\exp[-2i c_{11}Z^AZ^B]\cdot\\
[\Pi^A\otimes\Pi^B][R^A\otimes R^B]^\dag.
\end{multline}
For this construction, then, the single-qubit $J$-dependent phase shifts are removed and we acquire a fully entangling gate when
\be
\Phi^{AB}[
 \Lambda_{+}^A\Lambda_{+}^B
+\Lambda_{-}^A\Lambda_{-}^B
-\Lambda_{+}^A\Lambda_{-}^B
-\Lambda_{-}^A\Lambda_{+}^B]=3\pi.
\ee

So far, we have considered adiabatic gates for only three-spin encodings.  As there is some cost of assuring gauge freedom in this case, it may be worth quickly addressing whether this may be avoided using the 4-spin decoherence free subsystem, which lacks gauge.  Certainly, biases may be introduced on four spins to lift degeneracies and operated similarly to RX and AEON qubits; one example construction for this has recently been pointed out\cite{Sala2017}.  However, if we employ the Wigner-Eckart formula for inter-qubit exchange as above, we find that the first-order phase shifts for encoded states all vanish.  The reason is that encoded states combine $J_{ztnc}^A=0$ and $J_{ztnc}^B=0$ individual qubits into a total $J=0$ space, and therefore \refeq{exchangeformula} says that all inter-qubit matrix elements for encoded logical states are proportional to
$$
\sixj{J^A_{ztnc}}{1}{J^A_{ztnc}}{J^B_{ztnc}}{J}{J^B_{ztnc}}=
\sixj{0}{1}{0}{0}{0}{0} = 0.
$$
Interqubit couplings \emph{must} traverse through a leakage space, and therefore the lowest order phase shifts for an adiabatic gate are of the same order as leakage errors.  Alternatively, biases could be arranged to support degeneracies between leaked spaces; if the fourth spin is not coupled, for example, the problem may reduce to that of the $J$-invariant adiabatic gate as we have discussed, including all of its overhead and error considerations.

Thus far we have considered ideal pulsing.  However, there are several forms of errors in this gate, most dominantly:
\begin{itemize}
\item We have considered the first-order perturbative phase shift.  Unfortunately, higher-order terms do not obey the simple integer ratio between $J$-subspaces as the first order term, and therefore lead to errors.  These errors are suppressed by minimizing $(\Omega^{AB})/\min(\Omega^A,\Omega^B)$, which favors a slower coupling pulse.
\item We have considered the case of adiabatic pulsing.  Even if only the first-order energy shift is considered, rapid pulsing can drive unwanted transitions, including population transfer to leakage states.  These errors are suppressed with slower pulses and shaping of pulses.
\item Charge noise leads to random fluctuations in exchange amplitudes, including that of the biases $\Omega_k^{A,B}$.  Fluctuations in exchange rates are likely to be the errors limiting fidelity for this gate in existing devices, and typically such errors are suppressed with faster pulses.
\item Spurious inhomogeneous magnetic fields from nuclear spins and other sources lead to violations of the symmetries employed to construct this gate.  These error sources are suppressed at high enough values of $\Omega_k^{A,B}$, where charge noise is worse, and are otherwise minimized with shorter gate times.
\end{itemize}
We address these errors more quantitatively in \refsec{sec:errors}, but first we will establish an example construction in the next section.

\section{Example Construction}
\label{sec:examples}
So far, our presentation has been general in two ways, allowing for arbitrary choices of bias fields $\Omega^X_k$ and arbitrary selection of spins participating in the cross-DFS coupling exchange.  In this section, we will establish specific configuration for both sets of parameters, which will form the model for analyzing noise in \refsec{sec:errors}.

The single-qubit DFS biases will follow the ``linear'' arrangement for RX\cite{Medford2013} and AEON\cite{Shim2016} qubits, where $\Omega^A_{zt}=\Omega^A_{tn}=\Omega^A$, with $\Omega^A_{zn}=0$ and likewise $\Omega^B_{nt}=\Omega^B_{tn}=\Omega^B$ with $\Omega^B_{zn}=0$.  We will enforce non-degeneracy by assuming $\Omega^A \ne \Omega^B$, and for the numeric examples considered below we will arbitrarily set $\Omega^B = 1.7\Omega_A$.  Note that these exchange coupling terms, if understood via a Fermi-Hubbard model, are functions of the detunings and tunnel couplings of that model, and may be chosen to reduce some noise-sensitivities within those models.  We refer the reader to other presentations~\cite{Medford2013,Shim2016,Wardrop2016} to motivate those choices theoretically.

Unlike the RX and AEON cases, we will define our qubit in the DFS basis, i.e using the $J_{zt}^A$ quantum number.  As defined in \refsec{sec:DFSdef}, $W$ is the identity matrix. In this basis, $H_1^A$ is not diagonal; that is, the RX and AEON qubit states are not the same as the DFS qubit states defined by $J_{zt}^A$.  They are related by a simple SWAP of spins $t$ and $n$, which is handled in the formalism above the by the single-qubit rotation
\be
R^A = e^{-i\pi \vec{S}_t^A\cdot\vec{S}_n^A}\rightarrow
e^{-i\pi/4}
\begin{pmatrix}
1/2 & \sqrt{3}/2 & 0\\
\sqrt{3}/2 & -1/2 & 0 \\
 0         &  0   & 1\\
\end{pmatrix}
\ee
in the DFS basis.  This $\pi$ rotation may be done physically via a tuned pairwise exchange between spins $t$ and $n$.  The eigenstates of $H_1^A$ in the logically encoded subspace are then, in terms of $\ket{J_{zt}^A,J_{ztn}^A,m}$
DFS states,
$\ket{\xi_+^A,m} = V\ket{0,1/2,m} = R^A\ket{0;m}$
and
$\ket{\xi_-^A,m} = V\ket{0,1/2,m} = R^A\ket{1;m}$,
with respective energies $\xi_\pm^A = -\Omega\pm\Omega/2$ relative to the leakage state $\ket{Q}=\ket{1,3/2,m}$.  This rotation $R^A$ therefore provides both the conversion to $H_1^A$ eigenvectors notated $V$ in \refsec{sec:general_adiabatic} as well as the rotation that converts the diagonalized adiabatic gate action into qubit action.

For our refocussing pulse, we consider one (non-unique) choice for a 4-pulse composite DFS rotation,
\begin{multline}
\Pi^A = e^{-i\pi \vec{S}_z^A\cdot\vec{S}_z^A}
        e^{-i(\pi-\tan^{-1}\sqrt{8}) \vec{S}_t^A\cdot\vec{S}_n^A}\cdot\\
        e^{-i\tan^{-1}\sqrt{8} \vec{S}_z^A\cdot\vec{S}_t^A}
        e^{-i(\pi-\tan^{-1}\sqrt{8}) \vec{S}_t^A\cdot\vec{S}_n^A},
\end{multline}
which satisfies \refeq{Pidef} up to an overall phase with $\eta = (3\pi-\tan^{-1}\sqrt{8})/4$ and $\eta'=\pm\pi/2$, with sign depending on basis, corresponding to an encoded $Y$-rotation by angle $\pi$ in either the DFS qubit basis or the $\ket{\xi_\pm;m}$ RX/AEON basis.  Note that this is not a unique solution for the $Y$-pulse; e.g. Eng\cite{Eng2015} used a more symmetric pulse sequence for ease of calibration.

Our choice of cross-DFS coupling is to turn on exchange between the spins labeled $z$ in \refsec{sec:DFSdef}; this $zz$ coupling is shown in \reffig{pulse_schematic}, and was the ``Linear Geometry'' in Doherty and Wardrop\cite{Doherty2013}.  Calculation of the structure factors is a simple matrix computation following \refeq{structurecalc} and using $R_B$ for $V$.  In the case of $zz$ coupling, we obtain $\Lambda_+=0$ and $\Lambda_-=\sqrt{2/3}$,  giving
\be
c_{11}=\frac{(-1)^{J+1}}{4(J+2)!}\times \frac{2}{3}.
\ee
For $J=1$, this works out to $1/36$, as in Doherty and Wardrop\cite{Doherty2013} for the ``Linear Geometry".  The same result is had for $nn$, $zn$, or $nz$ couplings. If in contrast we couple $t$ spins, flipping the signs of off-diagonal elements in \refeq{matrixcalc}, we obtain $\Lambda_+^A = \sqrt{3/2}$ and $\Lambda_-^A = -1/\sqrt{6}$, yielding a result four times larger, again consistent for $J=1$ with Doherty and Wardrop's ``Butterfly Geometry."  Other geometries may be simply evaluated using these coefficients.  Critically, however, we find as promised that the $J=0$ subspace has in all cases a factor $-3$ faster amplitude.  The factor $-3$ ensures that it is always possible to produce compatible entangling gates in both $J$ subspaces, at least to first order in perturbation.

These example numbers are provided to help unpack the general notation of the preceding section. We re-emphasize that one does not require highly symmetric exchange couplings in the individual qubits for the two-qubit gate to succeed; this was employed only for simplicity.  In all cases, however, we find that the nonlinear phase $c_{11}$ is always proportional to $(-1)^{J+1}/(2+J)!$, enabling the $J$-invariant composite pulse sequence highlighted above.

\section{Errors}
\label{sec:errors}
\begin{figure*}
\includegraphics[width=\textwidth]{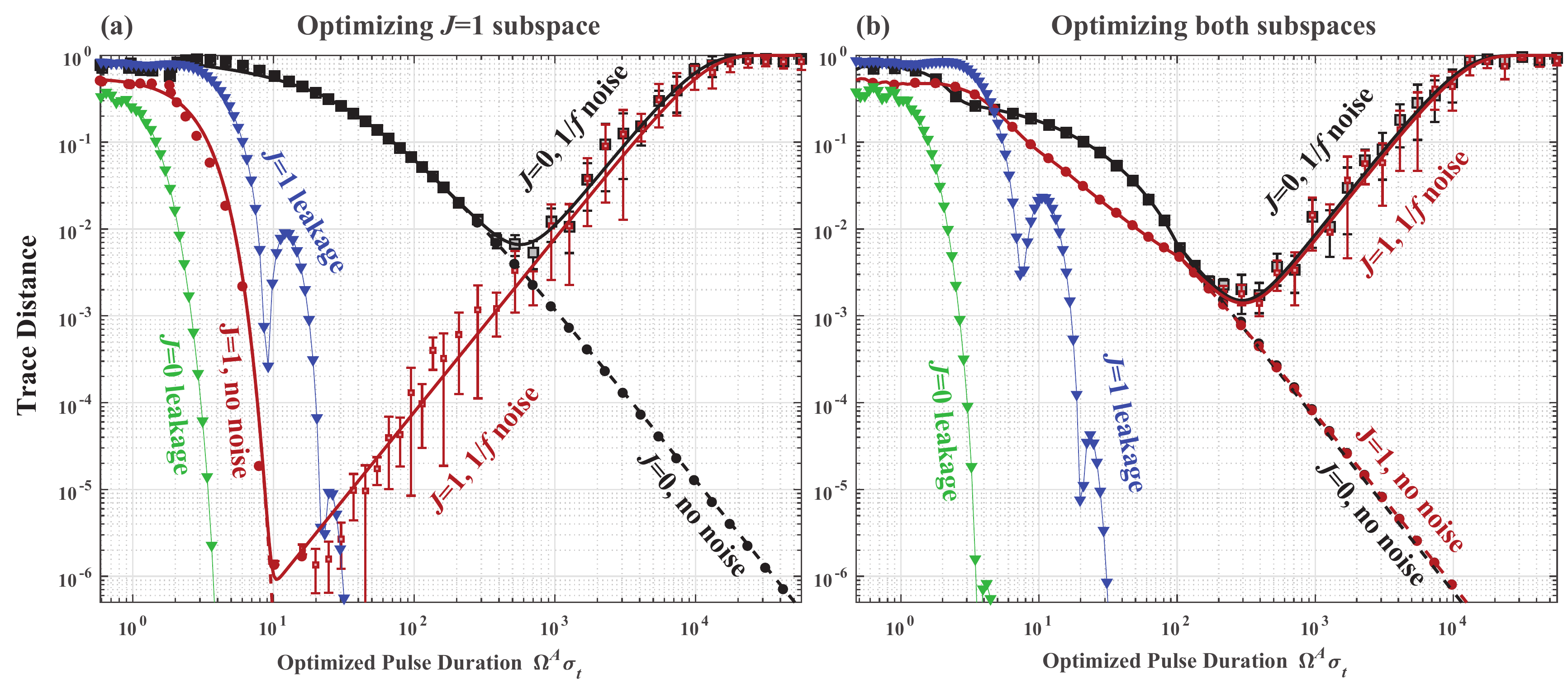}
\caption{Gate performance vs. pulsing time. In both plots, the pulse is Gaussian with amplitude $\Omega^{AB}$ that is optimized for a given pulsing time $\sigma_t$ for a trace-distance objective defined in \refeq{trace_distance}. The black (red) circles show the distance between the gate result and CZ within the $J=0$ ($J=1$) subspace, while the black (red) squares show the distance to CZ including  exchange-proportional charge noise with spectral density $S(f)=10^{-8}/f$.  The green (blue) triangles show the nonunitarity of the gate within the $J=0$ ($J=1$) subspace. (a) The pulse amplitude is optimized for a CZ in the $J=1$ subspace only.  Dashed black line shows phenomenological fit function $0.6\{1-\exp[1/(\Omega^A\sigma_t/22+(\Omega^A\sigma_t/45)^2)]\}$.
(b) The pulse amplitude is optimized to jointly make a CZ in both subspaces, placing three times more weight on the $J=1$ subspace.
In this case the dashed lines show data fits going as $\sim\exp[-(8/\Omega^A\sigma_t)^2]$; the solid black and red lines combine this with a fit going as $1-\exp(-1.1(\mathcal{N}\Omega^A\sigma_t/\mathcal{I})^2)$.
}
\label{errorfig}
\end{figure*}
We now consider some of the error sources on the $J$-invariant adiabatic CZ gate, using the exchange biases of the previous section and considering the $zz$ or ``linear" coupling case.  We focus on the effects of a finite pulse in the presence of charge noise, which include higher-order terms leading to different evolution in the $J$ subspaces, population transfer to leakage states, and charge noise affecting exchange-pulse accuracy.

We first consider the case of higher-order terms affecting both gauge-invariance and adiabaticity.  To evaluate performance, we presume the coupling pulse is Gaussian with root-mean-square width $\sigma_t$, i.e. $H_2^{AB}(t) = \Omega^{AB} \exp[-t^2/(2\sigma_t^2)]\vec{S}^A\cdot\vec{S}^B$.  We divide the pulse into piecewise constant intervals according to desired accuracy and exponentiate in the SU(5)$\times$SU(9) block-diagonal basis, unless considering magnetic noise in which case we employ all 20 states featuring $m=0$.  The coupling amplitude $\Omega^{AB}$ and the pulse width $\sigma_t$ are inversely related by the condition of setting the accumulated nonlinear phase to provide a fully entangling gate; for each simulation we find the optimal $\sigma_t$ for each $\Omega^{AB}$ to satisfy this condition.  There are several ways to optimize, giving quite different results.  Here we consider two cases.  Case 1, we optimize the amplitude $\Omega^{AB}$ as a function of $\sigma_t$ for the $J=1$ subblock only, relying on the fact the $J=0$ block will be close to ideal if the higher-order terms are small relative to first order.  Case 2, we optimize the pulse amplitude for both sublocks, putting three times more weight on the $J=1$ case due to there being three gauge configurations in $J=1$ compared to the single $J=0$ subspace.  We presume for what follows that the single qubit rotations $R^{A,B}$ and $\Pi^{A,B}$ are performed perfectly and instantaneously in order to focus on the two-qubit entangling gate.

To evaluate the gate fidelity, we examine two metrics.   First, we consider the unitarity of the encoded gate.  Notate the encoded 4$\times$4 logical subblock of the simulated gate as $\tilde{U}$, and recognize that $\tilde{U}^\dag\tilde{U}$ is not identity, due to the possibility of leakage.  Our first metric, then, is the trace distance between $\tilde{U}^\dag\tilde{U}$ and the identity, i.e. ``leakage" $= 1-|\text{Tr}\{\tilde{U}^\dag\tilde{U}\}|^2/16$.  Simulated values of leakage are shown as the green and blue triangles in \reffig{errorfig}, and as expected for adiabatic evolution with a Gaussian pulse, these drop roughly as Gaussian with respect to $\Omega^A\sigma_t$, representing a negligible contribution to error for $\sigma_t > 30/\Omega^A$.  For example, it is reasonable to set the bias exchange values at 10 Grad/sec, in which case pulsewidths of 3 nanoseconds suffice to suppress leakage.

Let us now consider errors due to higher order corrections in adiabatic phase accumulation.  For this, we consider the trace distance $D(\tilde{U},\ts{U}{target})$ between the simulated encoded evolution $\tilde{U}$ and the target encoded CZ gate, which for our construction has the form $\ts{U}{target}=e^{i\phi-i\pi ZZ/4}$; in particular we use
\be
 D = \frac{1}{2}+\frac{1}{32}\left|\Tr{\tilde{U}^\dag\tilde{U}}\right|^2
        -\frac{1}{16}\left|\Tr{\ts{U}{target}^\dag \tilde{U}}\right|^2.
        \label{trace_distance}
\ee
Figure~\ref{errorfig} shows $D$ as a function of $\Omega^A\sigma_t$ using black and red circles for $J = 0$ and $1$, respectively.  If optimizing the $J=1$ subblock, higher order contributions to phase are incorporated in the optimization, and so error due to these contributions is negligible, and $D$ is dominated by leakage errors, again enabling high fidelity gates within the $J=1$ subblock at relatively fast pulsing speeds.  However, the unoptimized $J=0$ subblock is worse, since higher-order terms are not an odd integer multiple of those in the $J=1$ subblock, as we have shown is the case for first-order terms.  As a result $D$ in the $J=0$ block is much worse for fast pulsing, and only reaches high fidelity operation in a limit where first-order perturbation theory is strongly valid, i.e. when $\sigma_t > 10^3/\Omega^a$, likely corresponding to hundreds of nanoseconds.

Unfortunately, long pulses must endure the detrimental effects of charge and magnetic noise.  For present silicon qubits, charge noise will be the more severe problem for this gate.  We note, as has Doherty and Wardrop\cite{Wardrop2016}, that if charge noise fluctuations were limited to very low frequencies, our refocusing single-qubit pulses would eliminate its effects.  However, charge noise typically presents itself as $1/f$, for which refocusing is less effective.

To model charge noise, we use a gate-referencing model, which we now explain.  We presume each exchange rate $\Omega^\alpha_{k}[\tilde{V}(t)]$ is a function of a noisy time dependent voltage $\tilde{V}(t)=V(t)+\delta V(t)$. Each applied exchange therefore becomes perturbed under charge noise to $\Omega^\alpha_k(V(t))+(d\Omega^\alpha_k/dV) \delta V(t)$.  We then assume $\delta V(t)$ is Gaussian noise drawn from power spectral density $S_{\delta V}(f)=\mathcal{N}^2/f$, which has noise amplitude $\mathcal{N}$ measured in volts.  We further define the insensitivity $\mathcal{I}$ as $\Omega^\alpha_k/|(d\Omega^\alpha_k/dV)|$, also measured in volts.  For more details of this model and its use in characterizing charge noise in a silicon triple dot, see Ref.~\onlinecite{Reed2016}.  In general, $\mathcal{I}$ will be a function of $\Omega^\alpha_k$, in particular if exchange is modulated via dot-to-dot detuning.  However, if employing barrier modulation, $\Omega^\alpha_k$ is roughly exponential in barrier-gate voltage, and $\mathcal{I}$ may be considered a constant. As shown in Ref.~\onlinecite{Reed2016}, this assumption is violated at low values of exchange, where a more complicated, disorder-influenced dependence is seen, and at high values of exchange, when dot pairs become highly merged.  However, for the present simulations we will fix the charge noise to a constant ratio $\mathcal{N}/\mathcal{I}=10^{-4}$, and we note that this is close to, but somewhat exceeding, state-of-the-art for silicon qubits.  In present silicon qubit devices $\mathcal{N}$ is of the order of tens of $\mu$V\cite{Eng2015,Freeman2016}, while $\mathcal{I}$ may be of order of 10s of mV\cite{Reed2016}.

Employing this model for $1/f$ noise at this optimistic but reasonable charge noise level, we simulate our composite gate multiple times for multiple instances of time-fluctuating noise, with the average values of $D$ plotted as squares in \reffig{errorfig} and the standard deviation as error bars.  The error tracks well the error expected just from applying the bias fields in $H_1^A$ and the echo pulse $\Pi^A$; as in the $Y$-echo experiment\cite{Eng2015}.  This in turn matches the integral over the spin-echo filter function, which scales as $(\mathcal{N}/\mathcal{I})\sigma_t^2$.  Since higher-order phase-correction errors scale as $\sigma_t^{-2}$, a minimum of $D$ is found.  This minimum is lower when simultaneously optimizing into both the $J=0$ and $J=1$ subspaces, as in \reffig{errorfig}b.  In this case, we find the minimum to be at $\sigma_t\approx 5\sqrt{\mathcal{I}/\mathcal{N}}/\Omega^A$, where the noise is approximately $100\mathcal{N}/\mathcal{I}$.  This emphasizes that charge noise improvements would be critical for a high-fidelity two qubit gate; improvements in $\mathcal{N}$ may result from better materials, while improvements in $\mathcal{I}$ may come from engineering operational sweetspots, as in Ref.~\onlinecite{Reed2016} or the RX\cite{Medford2013} and AEON\cite{Shim2016} qubits.

Finally, we briefly address magnetic noise, which comes dominantly from slow fluctuations of nuclear magnetization in both GaAs and Si material.  For this noise source, our composite construction has similar error to that analyzed by Doherty and Wardrop\cite{Wardrop2016}, and we refer the reader to this discussion.  However, we will point out that this adiabatic gate has a clear advantage relative to pulsed sequences such as Fong-Wandzura\cite{Fong2011} in that, in the adiabatic case, the biases present lift degeneracies between all states, highly suppressing magnetic noise, whereas pairwise exchange sequences have ongoing magnetic dephasing between degenerate encoded and leakage states.  As a result, the infidelity of pairwise-pulsed sequences is close to $(T/T_2^*)^2,$ where $T$ is the total time, which may be appreciable for a pulse sequence of 20 pulses or more.  Using experimentally typical parameters of 10 nanosecond pulses with 10 nanosecond idle times between pulses for a 20 pulse sequence, for a total of about 400 nanoseconds, this infidelity is appreciable even for quasistatic noise at a level consistent with isotopically enhanced silicon devices\cite{Eng2015,Reed2016}.  In contrast, we calculate the magnetic noise contribution to infidelity for an adiabatic gate of about the same duration to be an order of magnitude less.  A key difference, however, is that the ability to reduce magnetic noise in sequentially pulsed sequences is principally an engineering question of how fast one may apply well-calibrated pulses.  In contrast, as we have discussed, the duration of an adiabatic gate is lower-bounded by the need to suppress errors from high-order phase corrections and leakage.

\section{Conclusion}
Implementing entangling gates for exchange-only spin qubits is challenging because any cross-DFS exchange interaction can also cause leakage.  In pulsed schemes\cite{Fong2011,Zeuch2016}, the state of the system extends into leakage states in the middle of the pulse sequence, only returning to the logical subspace after all pulses are applied.  Doherty and Wardrop\cite{Doherty2013} provide an alternative approach where leakage is suppressed at all times using large exchange biases to raise the energy level of leaked states, and a cross-DFS pulse induces a perturbative entangling phase gate.  Although leakage still occurs to some degree, proper shaping of an adiabatic pulse strongly suppresses leakage error as a function of pulse time.  Importantly, those authors only considered an entangling a gate in the $J=1$ subspace, and we have shown that a modest change (two entangling pulses separated by single-qubit $Y$-echo pulses) can extend their proposal to $J=0$ as well, resulting in a $J$-invariant adiabatic CZ gate.  However, there are further consequences to implementing the gate in both $J$ gauges.  Whereas Doherty and Wardrop note that the amplitude of the entangling pulse can be calibrated to account for higher-order terms in perturbation~\cite{Doherty2013,Wardrop2016}, we have not found it to be possible to calibrate precisely for both $J$ subspaces simultaneously, though the result error does decay with pulse duration.  If gauges are prepared randomly (i.e. the gauge spin is loaded in a mixed state), then in general the applied gate will need to deal with both gauge subspaces.  Beyond gauge considerations, we have also shown that charge noise is problematic for this gate because of the substantial accumulated rotations on the single-qubit exchanges.  Even with the secondary benefit of the $Y$-echo pulses refocusing low-frequency charge noise~\cite{Wardrop2016}, the composite gate will be sensitive to noise at frequencies above the inverse of the entangling pulse duration, and this introduces an error source that increases with pulse duration.  The competition between the $J$-dependent higher-order perturbation terms and the errors due to charge noise results in an optimal pulse duration for a given set of device parameters.

As a final note, our construction employed only a single decoupling pulse.  Multipulse decoupling can in principle provide improved robustness against charge-noise, but of course the model we have employed in the present work treats the decoupling pulses as ideal and instantaneous.  As the decoupling pulses themselves will be noisy and necessarily applied at a finite rate, decoupling can extend only so far.   More analysis will be required to identify the ultimate limit, and therefore the practical utility, of adiabatic exchange-only two-qubit gates.

\bibliography{J_invariance}

\end{document}